\documentclass[3p]{elsarticle}
\usepackage[latin2]{inputenc}
\usepackage[english]{babel}
\usepackage{amsmath}
\usepackage{amssymb}
\usepackage{tabularx}

\newtheorem{thm}{Theorem}

\newtheorem{corollary}[thm]{Corollary}
\newtheorem{proposition}[thm]{Proposition}

\newtheorem{definition}[thm]{Definition}
\newproof{pf}{Proof}

\newcommand{\probdef}[1]{\begin{center}\noindent\fbox{\parbox{0.67\linewidth}{#1}}
\end{center}}

\DeclareMathAlphabet{\mathcal}{OMS}{cmsy}{m}{n}
\renewcommand{\S}{{\mathcal S}}
\renewcommand{\H}{{\mathcal H}}
\begin{document}

\begin{frontmatter}
\title{Completely inapproximable monotone and antimonotone
  parameterized problems\footnote{A preliminary version of the paper appeared in the proceedings of the 25th Annual IEEE Conference on Computational Complexity (CCC 2010). Research  supported by the European Research Council (ERC)  grant 
``PARAMTIGHT: Parameterized complexity and the search for tight complexity results,'' reference 280152.}}


\author{{Dániel Marx}}

\address{
Computer and Automation Research Institute\\
Hungarian Academy of Sciences (MTA SZTAKI)\\
Budapest, Hungary\\
dmarx@cs.bme.hu}

\begin{abstract}
  We prove that weighted circuit satisfiability for monotone or
  antimonotone circuits has no fixed-parameter tractable approximation
  algorithm with {\em any} approximation ratio function $\rho$, unless
  $\textup{FPT}\neq W[1]$. In particular, not having such an
  fpt-approximation algorithm implies that these problems have no
  polynomial-time approximation algorithms with ratio $\rho(OPT)$ for
  any nontrivial function $\rho$.
\end{abstract}

\begin{keyword}
inapproximability \sep fixed-parameter tractability \sep circuits \sep circuit satisfiability
\end{keyword}
\end{frontmatter}

\section{Introduction}
Recently, there has been increased interest in investigating
approximation in the context of parameterized complexity
\cite{iwpec-cgg,DBLP:journals/algorithmica/CaiH10,downey-fellows-mccartin,marx-approx,DBLP:conf/coco/EickmeyerGG08}.
Recall that a {\em parameterization} of a problem is a polynomial-time
computable function that assigns an integer $k$ to each problem
instance $x$. An {\em fpt-algorithm} for a parameterized problem is an
algorithm with running time $f(k)\cdot |x|^{O(1)}$, where $k$ is the
parameter of the input instance $x$ and $f$ is an arbitrary computable
function. A decision problem is {\em fixed-parameter tractable} (FPT)
with parameter $k$ if it can be solved by an fpt-algorithm. The
parameter can be any well-defined function of the input instance $x$;
for example, the required number of vertices in the solution, or the
maximum degree of the input graph. The standard way of turning an
optimization problem into a decision problem is to add a value $k$ to
the input instance and ask if there is a solution with cost at most/at
least $k$. Taking this value $k$ appearing in the input as the
parameter is called the {\em standard parameterization} of the
optimization problem. For a large number of NP-hard optimization
problems, the standard parameterization is FPT, for example, this is
the case for \textsc{Minimum Vertex Cover}, \textsc{Longest Path},
\textsc{Directed Feedback Vertex Set}, and \textsc{Multiway Cut}.
Intuitively, these results show that the problems can be solved
efficiently if the optimum is small. On the other hand, the
W[1]-hardness of other problems, such as \textsc{Maximum Clique} and
\textsc{Minimum Dominating Set} gives evidence that these problems are
not fixed-parameter tractable, and probably there is no significantly
better algorithms than solving the problem in time $n^{O(k)}$ by brute
force.

FPT approximation algorithms were introduced by three independent
papers \cite{iwpec-cgg,downey-fellows-mccartin,DBLP:journals/algorithmica/CaiH10}, see also
the survey \cite{marx-approx}. We follow here the notation of
\cite{iwpec-cgg}. {\em An fpt approximation algorithm with ratio
  $\rho$} for a minimization problem $P$ is an fpt-algorithm that,
given an instance $x$ of $P$ and a positive integer $k$, computes a
solution of cost at most $k\cdot \rho(k)$ if a solution of cost at
most $k$ exists; if there is no solution of cost at most $k$, then the
output can be arbitrary. The definition can be adapted to maximization
problems.
Note that the approximation ratio $\rho$ is a function of $k$, not the
input size: intuitively, if $k$ is small, then $k\cdot \rho(k)$ can be
still considered small. We say that problem is {\em fpt-approximable}
if it has an fpt approximation algorithm for some function $\rho$. As
we are proving hardness results in this paper, it will be convenient
to prove hardness result for a decision problem associated with
approximation. Following \cite{iwpec-cgg}, we say that an algorithm is
an {\em fpt cost approximation} with ratio $\rho$ if it distinguishes
(in fpt-time) between instances of optimum value at most $k$ and more
than $k\cdot \rho(k)$ (see Definition~\ref{def:cost}). It is clear
that it is sufficient to prove hardness results for this decision
variant to rule out the possibility of fpt approximation with ratio $\rho$.

On the positive side, there are a couple of nontrivial fpt
approximation algorithms. Seymour et al.~\cite{MR1433641} proved a
relation between the packing and the covering number of directed cycles,
which was subsequently made algorithmic by Grohe and
Grüber~\cite{DBLP:conf/icalp/GroheG07} in order to obtain an fpt
$\rho$-approximation algorithm for \textsc{Maximum Disjoint Cycles} in
directed graphs, with some unspecified function $\rho$. Marx and
Razgon~\cite{marx-multicut-ipl} gave an fpt 2-approximation algorithm
for \textsc{Edge Multicut}. However, later it was shown that
\textsc{Edge Multicut} (as well as \textsc{Vertex Multicut}) is
actually fixed-parameter tractable
\cite{marx-razgon-multicut,bousquet-multicut}.  Fellows (unpublished
result) showed that \textsc{Topological Bandwidth} has an fpt
approximation algorithm with ratio $k$ (see also
\cite{marx-approx}). On the negative side, it is known that
\textsc{Weighted Circuit Satisfiability} \cite{iwpec-cgg} and
\textsc{Independent Dominating Set} \cite{MR2466795} are not
fpt-approximable for {\em any} function $\rho$ (under standard
complexity assumptions).  However, these results are not very
enlightening as they use in an essential way that the considered
problems are not (anti)monotone: it is not true that every subset or
every superset of a solution is a solution. Therefore, it is very well
possible that every feasible solution of an instance is of the same
size, in which case any approximation algorithm has to actually find
an optimum solution. We call a minimization problem {\em monotone} if
any superset of a solution is also a solution.  Similarly, a
maximization problem is {\em antimonotone,} if any subset of the
solution is also a solution.  Inapproximability is usually a more
meaningful question for such problems.

The first inapproximability result in the fpt sense for a monotone problem was obtained
earlier and independently of the study of fpt approximation. As a key
step in showing that resolution is not automatizable, Alekhnovich and
Razborov \cite{DBLP:journals/siamcomp/AlekhnovichR08} showed that
there is no fpt 2-approximation algorithm for \textsc{Weighted
  Monotone Circuit Satisfiability}, unless every problem in the class W[P] can
be solved by a randomized fpt-algorithm. Eickmeyer et
al.~\cite{DBLP:conf/coco/EickmeyerGG08} improved this result in two
ways: they weakened the complexity assumption by removing the word
``randomized,'' and increased the ratio from 2 to any polylogarithmic
function. They conjectured that the problem has no fpt approximation
algorithm for any function $\rho$. Our first result confirms this
conjecture. The proof is completely different and much simpler than
the inapproximability proofs of \cite{DBLP:conf/coco/EickmeyerGG08,DBLP:journals/siamcomp/AlekhnovichR08}. Instead of using
expanders for gap amplification in a multi-layer circuit, our proof
achieves an arbitrary large gap using a simple construction based on
$k$-perfect hash functions. Furthermore, it is shown in 
\cite{DBLP:conf/coco/EickmeyerGG08} that appropriate gap-preserving
reductions can transfer the inapproximability result from
\textsc{Weighted Monotone Circuit Satisfiability} to other
parameterized minimization problems such as \textsc{Minimum Chain
  Reaction Closure}, \textsc{Minimum Generating Set}, and
\textsc{Minimum Linear Inequality Deletion}. The reductions transfer
our result as well, thus it follows that these problems are not
fpt-approximable either.

Eickmeyer et al.~\cite{DBLP:conf/coco/EickmeyerGG08} raised the
question if the maximization problem \textsc{Weighted Antimonote
  Circuit Satisfiability} is fpt-approximable; they conjectured that
this problem is also hard to approximate. Note that finding a maximum
weight solution for an antimonotone circuit is equivalent to finding a
minimum weight solution for a monotone circuit, but the
approximability of the two problems can be different. We prove this
conjecture by showing that the problem is not fpt-approximable, unless
$\text{FPT}\neq W[1]$.  The proof is somewhat similar to the
inapproximability results of
\cite{DBLP:journals/siamcomp/AlekhnovichR08,DBLP:conf/coco/EickmeyerGG08}
for the monotone version: it uses simple linear algebra (Reed-Solomon
codes) for error correction. However, the construction of the circuit
is much simpler, since we do not have to repeat the same circuit in
multiple layers to increase the gap.  In fact, we prove the result for
a special case of circuit satisfiability, which is a fairly natural
W[2]-complete combinatorial problem on hypergraphs: in the \textsc{Threshold Set}
problem, we are given a collection of subsets $\S$ of a universe $U$
with a weight $w(S)$ for each set $S\in \S$, the goal is to select the
maximum number of elements from $U$ such that every $S\in \S$ contains
at most $w(S)$ elements.

It would be interesting to obtain similar inapproximability results
for more restricted versions of circuit satisfiability and perhaps
even for natural combinatorial problems such as \textsc{Independent
  Set} and \textsc{Hitting Set}. The current paper is already a step
in this direction: we prove inapproximability results for
\textsc{Threshold Set} and for monotone/antimonotone circuit
satisfiability with certain bounds on the depth and weft of the
circuit. However, beyond a certain point, much deeper techniques would
be required than the elementary methods of the present paper. In
particular, the known proofs giving evidence that there is no
polynomial-time constant factor approximation algorithm for
\textsc{Hitting Set} and \textsc{Independent Set} all use the PCP
theorem. Thus ruling out fpt approximation for these problems would
require the use of (some generalization of) the PCP theorem.

Finally, let us mention that expressing the approximation ratio as a
function of the optimum (rather than as a function of the input size)
makes sense also in the context of polynomial-time approximation
algorithms. There are such results in the literature: for example,
Feige et al.~\cite{DBLP:journals/siamcomp/FeigeHL08} gave a polynomial-time $O(\sqrt{\log
  \text{OPT}})$ approximation for \textsc{Treewidth} and
Gupta~\cite{MR1974949} gave a polynomial-time $O(\text{OPT})$
approximation for \textsc{Directed Multicut}. However, there are no known
inapproximability results in this direction. For example, it is not
known whether there is a polynomial-time algorithm that, given a graph
with maximum clique size $k$, always finds a clique of size at least,
say, $\log \log \log k$. Showing that a certain problem is not
fpt-approximable would clearly imply that there is no polynomial-time
approximation algorithm with any ratio depending only on the optimum.
In particular, there are no such polynomial-time approximation
algorithms for the problems considered in this paper (under standard
assumptions).  Interestingly, the reverse implication also holds
\cite{DBLP:conf/icalp/GroheG07,marx-approx}: if a problem has an fpt
$\rho$-approximation algorithm for some function $\rho$, then there is
a polynomial-time approximation algorithm with approximation ratio
$\rho'(\textup{OPT})$ for some {\em other} function $\rho'$ (we sketch the
proof in Section~\ref{sec:preliminaries}). This means that if we want
to show for example that there is no polynomial-time algorithm finding
a clique of size $f(\textup{OPT})$ for {\em any} function $f$, then we are also
showing that the problem is not fpt-approximable. Therefore, it seems
that ruling out such polynomial-time algorithms is essentially a
problem belonging to parameterized complexity and requires the
understanding of fpt-approximability.

\section{Preliminaries}\label{sec:preliminaries}

\textbf{Parameterized approximation.}  An NP-optimization problem is
described by a tuple $(I,\textup{sol},\textup{cost},\textup{goal})$,
where $I$ is the set of instances, $\textup{sol}(x)$ is the set of
feasible solutions for instance $x$, the positive integer
$\textup{cost}(x,y)$ is the cost of solution $y$ for instance $x$, and
$\textup{goal}$ is either min or max. We assume that
$\textup{cost}(x,y)$ can be computed in polynomial time, $y\in
\textup{sol}(x)$ can be decided in polynomial time, and
$|y|=|x|^{O(1)}$ holds for every such $y$. We follow the notation Chen
et al.~\cite{iwpec-cgg} for the definitions of fpt-approximation:
\begin{definition}\label{def:standard-apx}
Let $O=(I,\textup{sol},\textup{cost},\textup{goal})$ be an
optimization problem (over some alphabet $\Sigma$) and let
$\rho:\mathbb{N}\to \mathbb{R}_{\ge 1}$ be a computable function such that $\rho(k)\ge 1$ for every
$k\ge 1$ and
\[
\begin{cases}
\text{$k\cdot \rho(k)$ is nondecreasing}  & \text{ if $\textup{goal}=\textup{min}$,}\\
\text{$k/\rho(k)$ is unbounded and nondecreasing} & \text{ if $\textup{goal}=\textup{max}$.}\\
\end{cases}
\]
An {\em fpt-approximation algorithm with approximation ratio $\rho$} for $O$
is an algorithm $\mathbb{A}$ that, given an input $(x,k)\in \Sigma^*\times
\mathbb{N}$ satisfying $\textup{sol}(x)\neq\emptyset$ and
\[
\begin{cases}
\textup{opt}(x)\le k & \text{ if $\textup{goal}=\textup{min}$,}\\
\textup{opt}(x)\ge k & \text{ if $\textup{goal}=\textup{max}$,}\\
\end{cases}\tag{*}
\]
computes a $y\in\text{sol}(x)$  such that
\[
\begin{cases}
\textup{cost}(x,y)\le k\cdot \rho(k) & \text{ if $\textup{goal}=\textup{min}$,}\\
\textup{cost}(x,y)\ge k/\rho(k)  & \text{ if $\textup{goal}=\textup{max}$.}\\
\end{cases}
\]
For inputs not satisfying condition (*), the output can be arbitrary;
in particular, this is the case if there is no solution for $x$.
Furthermore, the running time of $\mathbb{A}$ on input $(x,k)$ is $f(k)\cdot
|x|^{O(1)}$ for some computable function $f$.
\end{definition}
As mentioned in the introduction, an fpt $\rho$-approximation
algorithm implies that there is polynomial-time approximation
algorithm with ratio $\rho'(\text{OPT})$ for some function $\rho'$
(see \cite{DBLP:conf/icalp/GroheG07,marx-approx}). We sketch the proof
here for minimization problems, the proof for maximization problems is
analogous.  The proof requires the technical condition that we can
always find a trivial feasible solution (if exists) in polynomial
time.  Note that we defined $\textup{cost}(x,y)$ (and hence
$\textup{opt}(x)$) as a positive integer. Therefore, if some
maximization problem $O$ has an fpt $\rho$-approximation for some
function $\rho$, then running the algorithm with $k=1$ produces a
feasible solution in polynomial time (if it exists).  However, for minimization
problems, the existence of an fpt $\rho$-approximation algorithm does
not imply that it is always possible to find a feasible solution in
polynomial time.

\begin{thm}
  Let $O$ be a minimization problem such that a feasible solution can
  be found in polynomial time (if it exists). If $O$ has an fpt
  $\rho$-approximation algorithm $\mathbb{A}$ for some function
  $\rho$, then there is a polynomial-time algorithm $\mathbb{B}$ and a
  nondecreasing function $\rho'$ such that algorithm $\mathbb{B}$,
  given an instance $x$ of $O$ with $\textup{sol}(x)\neq\emptyset$, outputs a solution $y$ of $x$ with
  $\textup{cost}(x,y)\le \textup{opt}(x)\rho'(\textup{opt}(x))$.
\end{thm}
\begin{pf}
  Suppose that the running time of $\mathbb{A}$ can be bounded by
  $f(k)|x|^c$ for some function $f$ and constant $c$. Given an
  instance $x$, algorithm $\mathbb{B}$ does the following. First, it
  finds a feasible solution $y_x$ of $x$ in polynomial time. Then for
  every $i=1,\dots, n$, algorithm $\mathbb{B}$ simulates algorithm
  $\mathbb{A}$ with input $(x,i)$ for at most $|x|^{c+1}$ steps. If the
  simulation terminates within $|x|^{c+1}$ steps, then we check if the output
  is a feasible solution. Let $y$ be the best feasible solution among
  the at most $n$ outputs of the simulations and the feasible solution
  $y_x$. 

We claim $\textup{cost}(x,y)\le
  \textup{opt}(x)\rho'(\textup{opt}(x))$ for some function $\rho'$.
  Let $k:=\textup{opt}(x)$. If $f(k)\le n$ and $k\le n$, then the
  simulation of $\mathbb{A}$ on $(x,k)$ terminates in $f(k)|x|^c\le
  |x|^{c+1}$ steps with a solution of cost at most $k\cdot
  \rho(k)$, which means that $\textup{cost}(x,y)\le
  \textup{opt}(x)\rho(\textup{opt}(x))$. Otherwise, let $\tau(k)$ be the
  maximum of $\textup{cost}(x,y_x)/\textup{opt}(x)$, taken over all
  instances of size at most $\max\{f(k),k\}$. Note that this is well
  defined, as there are only a finite number of such
  instances. Therefore, if $n \le \max\{f(k),k\}$, then 
$\textup{cost}(x,y)\le \textup{cost}(x,y_x) \le 
  \textup{opt}(x)\tau(\textup{opt}(x))$. Thus the function
  $\rho'(k)=\max\{\rho(k),\tau(k)\}$ satisfies the requirements.
\qed\end{pf}

Chen~et al.~\cite{iwpec-cgg} defined a weaker notion of
approximability, which is a decision algorithm solving the gap version
of the decision problem associated with the optimization
problem. Similarly to \cite{DBLP:conf/coco/EickmeyerGG08}, we consider
only this weaker notion in our inapproximability results (thus in fact
making the results slightly stronger).

\begin{definition}\label{def:cost}
Let $O$ and $\rho$ be as in Definition~\ref{def:standard-apx}. A
decision algorithm $\mathbb{A}$ is an {\em fpt cost approximation
  algorithm} for $O$ with {\em approximation ratio $\rho$} if for
every input $(x,k)\in \Sigma^*\times 
\mathbb{N}$ with $\textup{sol}(x)\neq\emptyset$, its output satisfies the following conditions:
\begin{enumerate}
\item If $\begin{cases}
\text{$k< \textup{opt}(x)$ and $\textup{goal}=\textup{min},$}\\
\text{$k> \textup{opt}(x)$ and $\textup{goal}=\textup{max},$}\\
\end{cases}
$\\
then $\mathbb{A}$ rejects $(x,k)$.
\item If $\begin{cases}
\text{$k\ge \textup{opt}(x)\cdot \rho(\textup{opt}(x))$ and $\textup{goal}=\textup{min},$}\\
\text{$k\le \textup{opt}(x)/\rho(\textup{opt}(x))$ and $\textup{goal}=\textup{max},$}\\
\end{cases}
$\\
then $\mathbb{A}$ accepts $(x,k)$.
\end{enumerate}
Furthermore, the running time of $\mathbb{A}$ on input $(x,k)$ is $f(k)\cdot
|x|^{O(1)}$ for some computable function $f$.
\end{definition}
Clearly, an fpt approximation algorithm with ratio $\rho$ implies
that there is an fpt cost approximation algorithm with the same ratio.

\textbf{Circuits.} A {\em Boolean circuit} is a directed acyclic
graph, where each node with indegree $>1$ is labeled as either an AND
node or as an OR node, each node of indegree 1 is labeled as a
negation node, and each node of indegree 0 is an input node.
Furthermore, there is a node with outdegree 0 that is the
output node. Given an assignment $a$ from the input nodes of circuit
$C$ to $\{0,1\}$, we say that assignment $a$ {\em satisfies} $C$ if
the value of the output node (computed in the obvious way) is 1. The
{\em weight} of an assignment is the number of input nodes with value
1. Circuit $C$ is {\em $k$-satisfiable} if there is a weight-$k$ assignment satisfying $C$. 

We denote by $|C|$ the number of nodes in the circuit.  The {\em
  depth} of circuit $C$ is the maximum length of a directed path from
an input node to the output node. The {\em weft} of a circuit is the
maximum number of large nodes  on any path from an input
node to the output, where ``large'' means that the node has indegree greater than 2.\footnote{We follow \cite{grohe-flum-param} in the definition of weft. In the original definition of Downey and Fellows \cite{MR2001b:68042}, ``large'' is defined as having indegree larger than a preagreed fixed bound.}
 Note that any circuit can be transformed into a
equivalent circuit of weft 0 by replacing each large node with a
sequence of nodes with indegree $2$. Thus bounding the weft is
meaningful only if we simultaneously bound the depth as well. The
notion of weft plays an important role in parameterized complexity and
in defining the classes of the W-hierarchy. For the definitions of the
classes $W[1]$, $W[P]$, etc., the reader is referred to standard texts
such as \cite{MR2001b:68042,grohe-flum-param}. Let us mention here
briefly that a parameterized problem $Q$ is in the class $W[t]$ if
there is a constant $d$ such that there is a parameterized reduction
from $Q$ to the satisfiability of circuits with depth $d$ and weft
$t$, while $Q$ is in $W[P]$ if there is a parameterized reduction from
$Q$ to the circuit satisfiability problem without any restriction on
depth and weft. The most important property of a parameterized
reduction is that it the parameter of the constructed instance is
bounded by a function of the original instance.

In the present paper, we investigate weft only to
see how restricted the classes of circuits are for which we manage to prove
inapproximability and to see what the exact parameterized
complexity assumptions are that we need for the results. If the reader is
not interested in these issues, then these discussions can be
ignored.

A circuit is {\em monotone} if it
contains no negation nodes. A circuit is {\em antimonotone} if the
unique inneighbor of each negation node is an input node, and every
outneighbor of an input node is a negation node. We define the
following two optimization problems:
\medskip

\probdef{
\textsc{Monotone Circuit Satisfiability}

\begin{tabular}{rl}
{\em Input:} & A monotone circuit $C$\\
{\em Solutions:} & All satisfying assignments $a$ of $C$\\
{\em Cost:} & The weight of satisfying assignment $a$\\
{\em Goal:} & min
\end{tabular}}
\medskip

\probdef{
\textsc{Antimonotone Circuit Satisfiability}

\begin{tabular}{rl}
{\em Input:} & An antimonotone circuit $C$\\
{\em Solutions:} & All satisfying assignments $a$ of $C$\\
{\em Cost:} & The weight of satisfying assignment $a$\\
{\em Goal:} & max
\end{tabular}}
\medskip

It is known that standard parameterizations of \textsc{Monotone Circuit Satisfiability} and
\textsc{Antimonotone Circuit Satisfiability} are W[P]-complete \cite{MR2001b:68042,grohe-flum-param}.

\section{Monotone problems}

We prove our main result on monotone circuits in this section:
\begin{thm}\label{th:monotone}
\textsc{Monotone Circuit Satisfiability} is not fpt cost approximable, unless $\textup{FPT}=\textup{W[P]}$.
\end{thm}
\begin{pf}
  Suppose that there is an fpt cost approximation algorithm $\mathbb{A}$
  for \textsc{Monotone Circuit Satisfiability} with approximation
  ratio $\rho$. We show that this algorithm $\mathbb{A}$ can be used
  to solve the standard parameterization of \textsc{Monotone Circuit
    Satisfiability} in fpt-time, implying
  $\textup{FPT}=\textup{W[P]}$.

Let $C$ be a monotone circuit with $n$ inputs. There is a natural
correspondence between the assignments to the $n$ inputs and the
subsets of $[n]$ (as usual, $[n]$ denotes $\{1,\dots,n\}$). Thus we can interpret $C$ as a Boolean function
$C(S)$ defined over the subsets $S\subseteq [n]$. 

Let $\H$ be a family of functions from $[n]$ to $[k']$. We say that
$\H$ is a {\em $k'$-perfect family of hash functions} if for every
$k'$-element set $S\subseteq [n]$, there is an $h\in \H$ such that $h$
is one-to-one on $S$, i.e., $h(s)\neq h(s')$ for every $s,s'\in S$,
$s\neq s'$. By Alon et al.~\cite{MR1411787}, it is possible to construct a
$k'$-perfect family $\H$ of size $2^{O(k')}\log n$ in time that is
polynomial in $n$ and $|\H|$.

Let monotone circuit $C$ and integer $k$ be the input of a
\textsc{Monotone Circuit Satisfiability} instance. Let $k':=\lceil
\rho(k)\cdot k \rceil$ (recall that $\rho$ is computable) and let $\H$
be a $k'$-perfect family of hash functions from $[n]$ to $[k']$. We
define the following function:
\begin{equation*}\label{eq:fprime}
C'(S):=\bigwedge _{h\in \H}
\bigvee_{T\in \binom{[k']}{\le k}} C(S\cap h^{-1}(T)),
\end{equation*}
where $\binom{[k']}{\le k}$ denotes the subsets of $[k']$ of size at
most $k$ and $h^{-1}(T)=\{i\in [n]\mid h(i)\in T\}$.  It is clear that
we can construct a monotone circuit $C'$ expressing the function
$C'(S)$ in time $g(k)|C|^{O(1)}$. We claim that
\begin{enumerate}
\item[(1)] if $C$ is
$k$-satisfiable, then $C'$ is also $k$-satisfiable, and
\item[(2)] if $C$ is
not $k$-satisfiable, then $C'$ is not $k'$-satisfiable.
\end{enumerate}
Let us run $\mathbb{A}$ with input $(C',k')$; as the size of $C'$ is
$g(k)|C|^{O(1)}$ and $k'$ is a
function of $k$, the running time of $\mathbb{A}$ is $f(k)|C|^{O(1)}$ for some
function $f(k)$.  If $C$ is $k$-satisfiable, then $\textup{opt}(C')\le
k$ and $k'\ge k\cdot \rho(k) \ge \textup{opt}(C')\cdot
\rho(\textup{opt}(C'))$, thus $\mathbb{A}$ accepts. On the other hand,
if $C$ is not $k$-satisfiable, then $\textup{opt}(C')>k'$ and
$\mathbb{A}$ rejects.  This means that we can solve the
standard parameterization of \textsc{Monotone Circuit Satisfiability} using
algorithm $\mathbb{A}$, implying $\textup{FPT}=\textup{W[P]}$.

To prove (1), we show that any weight-$k$ satisfying assignment of $C$
satisfies $C'$ as well.  Let $S\subseteq [n]$ be a weight-$k$
satisfying assignment of $C(S)$. We have to show that the disjunction
in $C'$ is true for every $h\in \H$. Let $T:=\{h(s): s\in S\}$;
clearly, $|T|\le k$. By definition, $S\subseteq h^{-1}(T)$, thus
$C(S\cap h^{-1}(T))=C(S)=1$.  Thus the disjunction is satisfied
by the term corresponding to this $T$.

To prove (2), let $S$ be a weight-$k'$ satisfying assignment of $C'$.
Let $h\in \H$ be a hash function that maps $S$ one-to-one; since $\H$
is $k'$-perfect, at least one such function exits. We claim that the
disjunction corresponding to this $h$ is not satisfied. To see this,
observe that for every $T\in \binom{[k']}{\le k}$, we have $|S\cap
h^{-1}(T)|\le k$: for every $t\in T$, function $h$ maps at most one
element of $S$ to $t$. Thus if $C(S\cap h^{-1}(T))=1$ for some $T\in
\binom{[k']}{\le k}$, then $S\cap h^{-1}(T)$ is a satisfying
assignment of weight at most $k$ for $C$, a contradiction.
\qed\end{pf}

Inspection of the proof shows that if circuit $C$ has depth $d$ and
weft $w$, then we can construct $C'$ such that it has depth $d+2$ and
weft $w+2$. Since the W[2]-complete \textsc{Hitting Set} problem can be expressed as a monotone
circuit having depth $2$, we get the following version of
Theorem~\ref{th:monotone}:
\begin{corollary}
If \textsc{Monotone Circuit
  Satisfiability} is fpt cost approximable for circuits with depth $4$, then $\textup{FPT}=W[2]$.
\end{corollary}
Note that this corollary shows the inapproximability of a more
restricted problem, but the assumption is somewhat stronger than in
Theorem~\ref{th:monotone}.

The weft of a depth-4 circuit is clearly at most 4. We can decrease
the weft at the cost of increasing the depth as follows. First, the
disjunction in $C'$ can be implemented without increasing the weft by
using a composition of OR nodes with indegree two. This increases the
depth, but this increase is bounded by a function of $k$. Second,
instead of starting with a weft-2 circuit $C$, we can start with a weft-1
circuit:
\begin{proposition}
There is a function $d(k)$ such that the standard parameterization of \textsc{Monotone Circuit
  Satisfiability} is $W[1]$-hard for instances with weft 1 and depth
bounded by $d(k)$.
\end{proposition}
\begin{pf}
  We reduce from the W[1]-hard \textsc{Multicolored Clique} problem
  \cite{DBLP:journals/tcs/FellowsHRV09}: given a graph $G$ and a
  proper $k$-coloring of the vertices, find a clique that contains
  exactly one vertex from each color class. Let $V_1$, $\dots$, $V_k$ be the
  $k$ color classes. We construct a monotone circuit $C$ where the inputs
  $x_1,\dots,x_n$ correspond to the vertices of $G$ and $C$ expresses
  the following function:
\[
\bigwedge_{1 \le i < j \le k}\bigvee_{\substack{x_a\in V_i, x_b\in
    V_j\\\text{$x_a,x_b$ are adjacent}}} (x_a\wedge x_b).
\]
It is easy to see that every weight-$k$ satisfying assignment of $C$
corresponds to a $k$-clique of $G$. Note that in a weight-$k$
satisfying assignment, in each color class exactly one variable is 1.
If the first conjunction is implemented with AND nodes of indegree 2,
then the weft of $C$ is 1 and the depth is bounded by a function $k$.
\qed\end{pf}
Putting together, we obtain:
\begin{corollary}
  There is a function $d(k)$ such that if \textsc{Monotone Circuit
    Satisfiability} is fpt cost-approximable for instances with weft
  $2$ and depth at most $d(k)$, then $\textup{FPT}=W[1]$.
\end{corollary}

\section{Antimonotone problems}
The main result of this section is that \textsc{Antimonotone Circuit
  Satisfiability} is not fpt approximable. We prove the result by
showing inapproximability for the following combinatorial problem:
\medskip

\probdef{
\textsc{Threshold Set}

\begin{tabularx}{\linewidth}{rX} 
{\em Input:} & A collection ${\S}$ of subsets of $U$ with a positive
integer weight $w(S)$ for each set $S\in \S$.\\
{\em Solutions:}  & A set $T\subseteq U$ such that
$|T\cap S|\le w(S)$ for every $S\in \S$.\\
{\em Cost:} & $|T|$\\
{\em Goal:} & max.
\end{tabularx}
}
\medskip

It is not difficult to express \textsc{Threshold Set} as an
antimonotone circuit. In particular, we discuss at the end of this
section how to express the \textsc{Threshold Set} instances
constructed in the inapproximability proof. Let us mention that it can
be shown that \textsc{Threshold Set} is W[2]-complete, although we do
not need this fact here.

The reduction showing the inapproximability of \textsc{Threshold Set}
uses Reed-Solomon codes to construct instances and relies on the erasure
correction properties of such codes to argue that finding even an
approximate solution is as hard as finding an optimum solution.  Let
us recall some basic facts about Reed-Solomon codes (no more
background is required for understanding the current paper). Let $F_q$
be the $q$-element finite field. For some $k \le D < q$, the
Reed-Solomon code is a function $RS: F^k_q\to F^D_q$ defined as
follows. Let us pick arbitrary  distinct nonzero elements
$\alpha_1$, $\dots$, $\alpha_D$ from $F_q$. For every
$m=(m_1,\dots,m_k)\in F^k_q$ and $1\le i \le D$, we define $RS(m)$
such that its {$i$-th} component
is $RS(m)[i]:=\sum_{j=1}^{k}\alpha_i^j m_j$.
It is well known that the Reed-Solomon code can correct $D-k$
erasures, or in other words, the original vector $m$ can be recovered
from any $k$ components of  $RS(m)$. We can state this formally as
follows:
\begin{proposition}\label{prop:rs}
For every $a,b\in
F^k_q$ and $I\subseteq \{1,\dots,D\}$ with $|I|=k$, if
$RS(a)[i]=RS(b)[i]$ for every $i\in I$, then $a=b$.
\end{proposition}
\begin{pf}
Let $w_i\in F^k_q$ be a column vector whose $j$-th component is
$\alpha_i^j$, which means that $RS(m)[i]=m\cdot w_i$. Note that
the vectors $\{w_i\mid i\in I\}$ are linearly independent by the well-known
properties of Vandermonde matrices, hence they form a basis of
$F^k_q$. Therefore, if $a$ and $b$ have the same inner product with
every vector in this basis, then $a=b$.
\qed\end{pf}
Now we are ready to prove the inapproximability result for   \textsc{Threshold Set}:

\begin{thm}
  \textsc{Threshold Set} is not fpt cost approximable, unless
  $\textup{FPT}=W[1]$.
\end{thm}
\begin{pf}
  Suppose that there is an fpt cost approximation algorithm
  $\mathbb{A}$ for \textsc{Threshold Set} with approximation ratio
  $\rho$. We show how to solve \textsc{Maximum Clique} in fpt-time
  using $\mathbb{A}$. Let $G$ be a graph where we have to decide
  if a clique of size $k$ exists. Let $D$ be the smallest integer such
  that $D/\rho(D)\ge k$. Note that such a $D$ exists and can be
  computed in time depending only on $k$ (recall that $\rho$ is computable). Let $n$ be the number of
  vertices of $G$. By adding isolated vertices, we can assume that
  $n>D^k$ and $n=2^{k\ell}$ for some integer $\ell\ge 1$; these
  additional vertices can increase the size of the graph only by a
  factor of at most $D^k\cdot 2^{k}$, which is a function of $k$ only.
  Let $q=2^{\ell}$ and let $F_q$ be the $q$-element finite field. We
  identify the vertices of $G$ with $F^k_q$, i.e., we will consider
  vertices as $k$-dimensional row vectors over $F_q$. For a vector
  $g$, we denote by $g[i]$ its $i$-th component.

  We construct an instance of \textsc{Threshold Set} as follows. The
  set $U$ consists of $n\cdot D$ elements: $U=\{s_{d,g} : 1 \le d \le
  D, g\in F^k_q\}$. Let us define the collection $\S$ of subsets in the
  input. First, for every $1 \le d\le D$, $\S$ contains the set
  $S_d:=\{s_{d,g} : g\in F^k_q\}$. By setting $w(S_d)=1$ for every $1
  \le d \le D$, we ensure that the solution $T$ contains at most one
  element with first index $d$. Thus a solution can be interpreted as
  a collection of at most $D$ vectors from $F^k_q$. Equivalently, every
  solution can be interpreted as a table of size $k\times D$, where
  each entry is either empty or contains an element of $F_q$. More
  precisely, if the solution contains element $s_{d,g}$ of $S_d$, then
  the $d$-th column contains components of the $k$-dimensional vector
  $g\in F^k_q$; if the solution contains no element of $S_d$, then the
  $d$-th column of the table is empty.  This table will be interpreted
  as an encoding of the $k$ vertices of the clique: the
  $D$-dimensional vector in the $i$-th row is interpreted as the
  encoding of the $i$-th vertex of the clique by a Reed-Solomon code
  $F^k_q\to F^D_q$. (Note that the $k$-dimensional column vectors of
  this table could be also interpreted as vertices, but this is {\em
    not} what we are doing here.) By the properties of the
  Reed-Solomon code, any $k$ full columns already describe a clique of
  size $k$.  Note that $n= 2^{k\ell}$ and $D^k<n$ implies $D<2^\ell=q$, as
  required by the Reed-Solomon code.

In order to enforce this interpretation, we add further sets to $\S$ as follows. Let $(X,a,b,u,v)$
be such that
\begin{itemize}
\item $X$ is a $k$-element subset of $\{1,\dots,
  D\}$,
\item $1 \le a < b \le k$,
\item $u,v\in F^k_q$ are nonadjacent vertices of $G$ (including the
  possibility $u=v$).
\end{itemize}
For each such 5-tuple, we add the following set to $\S$:
\[
S_{X,a,b,u,v}=\{s_{j,g} : j\in X, g\in F^k_q, g[a]=RS(u)[j],
g[b]=RS(v)[j]\}.
\]
The weight of each such set is $k-1$. If we interpret a solution as a
table of size $k\times D$ (as described above), then weight $k-1$ of
the set $S_{X,a,b,u,v}$ ensures that it is not possible that the
entries in row $a$ and columns $X$ agree with $RS(u)$ and at the same
time the entries in row $b$ and columns $X$ agree with $RS(v)$. As we
want the $a$-th and $b$-th vertex of the clique to be adjacent, we
require this for every nonadjacent $u$ and $v$.  This completes the
construction of the \textsc{Threshold Set} instance $x$. Note that the
size of instance $x$ is $g(k)n^{O(1)}$ for some function $g(k)$
depending only on $k$.

We claim that
\begin{enumerate}
\item[(1)] if $G$ has a $k$-clique, then $x$ has a solution $T$ of size $D$ and
\item[(2)] if $G$ has no $k$-clique,
then every solution of $x$ has size less than $k$. 
\end{enumerate}
If these claims are true, then we can decide whether $G$ has a
$k$-clique by running $\mathbb{A}$ on $(x,k)$. If $G$ has a $k$-clique, then
we have $\textup{opt}(x)\ge D$, which means that
$\textup{opt}(x)/\rho(\textup{opt}(x))\ge D/\rho(D) \ge k$, and
$\mathbb{A}$ accepts. On the other hand, if there is no $k$-clique,
then $\textup{opt}(x)< k$, and $\mathbb{A}$ rejects.  As 
the size of $x$ is $g(k)n^{O(1)}$, the
running time of $\mathbb{A}$ can be bounded as $f(k)n^{O(1)}$ for
some computable function $f(k)$. It follows that the construction of
the \textsc{Threshold Set} instance and running algorithm $\mathbb{A}$
is an fpt-time algorithm for solving the \textsc{Maximum Clique}
problem, implying $\textup{FPT}=\textup{W[1]}$.

To prove (1), suppose that $G$ has a $k$-clique $v_1$, $\dots$, $v_k\in F^k_q$.
For every $1 \le j \le D$, we define the vector $g_j\in F^k_q$ such that
$g_j[i]=RS(v_i)[j]$ for every $1\le i \le k$. We claim that the
$D$-element set $T:=\{s_{j,g_j}: 1\le j \le D\}$ is a feasible
solution. It is clear that the edge $S_j$ contains exactly one element
of $T$. Let us verify that every set $S_{X,a,b,u,v}$ of $\S$ contains
at most $k-1$ elements of $T$. Since $u$ and $v$ are not adjacent, but 
$v_a$ and $v_b$ are adjacent, at least one of $v_a\neq u$ or $v_b\neq
v$ holds. Suppose that $v_a\neq u$ (the case $v_b\neq v$ is similar).
By Prop.~\ref{prop:rs}, there has to be a $j\in X$ such that
$RS(v_a)[j]\neq RS(u)[j]$. This implies that $s_{j,g_j}\in T$ is not in $S_{X,a,b,u,v}$:
we have $g_j[a]=RS(v_a)[j]$ by the definition of $g_j$, while
$s_{j,g}\in S_{X,a,b,u,v}$ only if $g[a]=RS(u)[j]$. Therefore,
$S_{X,a,b,u,v}\cap T$ contains no element from $S_j$, implying that
the intersection has size at most $k-1$.

To prove (2), suppose now that there is a solution $T$ of size at
least $k$. Define $J\subseteq \{1,\dots,D\}$ such that $j\in J$ if and
only if $S_j\cap T\neq \emptyset$. Since $|S_j\cap T|\le 1$ for every
$1\le j \le D$, we have $|J|\ge k$. Let $X$ be a $k$-element subset of
$J$. For every $j\in X$, there is a unique value $g_j$ such that
$s_{j,g_j}\in T$. For $1 \le i \le k$, let $v_i$ be the unique vertex
satisfying $RS(v_i)[j] =g_j[i]$ for every $j\in X$ (the existence
and the uniqueness of $v_i$ follows from Prop.~\ref{prop:rs}). We claim that $v_1$, $\dots$, $v_k$ form a
clique in $G$. Suppose that $v_a$ and $v_b$ are not adjacent. Then the
set $S_{X,a,b,v_a,v_b}$ is a member of $\S$. It is easy to see that
$v_{j,g_j}$ is in this set for every $j\in X$: by the definition of $v_a$
and $v_b$, we have $g_j[a]=RS(v_a)[j]$ and $g_j[b]=RS(v_b)[j]$ for
every $a,b\in X$.  However, the set $S_{X,a,b,v_a,v_b}$ has weight
$k-1<|X|$, a contradiction.
\qed\end{pf}

Let us discuss how the \textsc{Threshold Set} instances
constructed in the proof can be expressed as circuits. Let the inputs $s_{j,g}$
correspond to the elements of $U$. The requirement $|T\cap S_j|\le 1$
can be expressed by requiring $\bar s_{j,g'}\vee \bar s_{j,g''}$ for
every distinct $g',g''\in F^k_q$. The set $S_{X,a,b,v_a,v_b}$ (which has
weight $k-1$) intersects $S_j$ only if $j\in X$, that is, for $k$
values of $j$.  Therefore, $|T\cap S_{X,a,b,v_a,v_b}|\le k-1$ is true
if and only if there is a $j\in X$ such that $T$ and
$S_{X,a,b,u,v}\cap S_j$ are disjoint. This means that the
\textsc{Threshold Set} instance can be expressed as
\[
\left(\bigwedge_{j=1}^D\ \bigwedge_{\substack{g',g''\in F^t_q\\g'\neq g''}}(\bar
s_{j,g'}\vee \bar s_{j,g''})\right) \\\wedge \left(\bigwedge_{S_{X,a,b,u,v}\in
  \S}\ \bigvee_{j\in X}\ \bigwedge_{s_{j,g}\in S_{X,a,b,u,v}}\hspace{-1em}\bar s_{j,g}\right).
\]
We obtain that the instance can be implemented by an antimonotone
formula of depth 3. If the first conjunction and the disjunction over
$j\in X$ are implemented by nodes of indegree 2, then we can get
weft-2 circuit having depth bounded by a function of $k$.

\begin{corollary}
There is a function $d(k)$ such that if  \textsc{Antimonotone Circuit
  Satisfiability} is fpt cost
approximable for instances of depth $3$, or for instances of weft 2
and depth bounded by $d(k)$, then $\textup{FPT}=W[1]$. 
\end{corollary}

\section*{Acknowledgments}
I'm grateful to Martin Grohe for useful discussions and to the
anonymous referees for comments that improved the presentation of the
paper.

\bibliography{inapprox}
\bibliographystyle{abbrv}

\end{document}